\documentclass[twocolumn,showpacs,preprintnumbers,amsmath,amssymb]{revtex4-1}

\usepackage{dcolumn}
\usepackage{bm}

\usepackage{color}

\usepackage{graphicx}
\usepackage{natbib}
\usepackage{subfigure}
\usepackage{cases}
\usepackage{epstopdf}
\newcommand{\beq}{\begin{equation}}
\newcommand{\eeq}{\end{equation}}

\begin{document}
\def\bfB{\mbox{\bf B}}
\def\bfQ{\mbox{\bf Q}}
\def\bfD{\mbox{\bf D}}
\def\etal{\mbox{\it et al}}
\title{From Weakly to Strongly Magnetized Isotropic MHD Turbulence}
\author{Alexandros Alexakis}

\affiliation{Laboratoire de Physique Statistique de l'Ecole Normale
Sup\'erieure, UMR CNRS 8550, 24 Rue Lhomond, 75006 Paris Cedex 05, France.}

\date{\today}

\begin{abstract}
High Reynolds number isotropic magneto-hydro-dynamic turbulence in the presence of large scale magnetic fields is investigated as a function of the magnetic field strength.
For a variety of flow configurations the energy dissipation rate $\epsilon$ follows the Kolmogorov scaling $\epsilon\propto U_{rms}^3/\ell$ 
even when the large scale magnetic field energy is twenty times larger than the kinetic. Further increase of the magnetic energy showed
a transition to the $\epsilon \propto U_{rms}^2 B_{rms} /\ell$ scaling implying that magnetic shear becomes more efficient at this point at cascading the energy 
than the velocity fluctuations. Strongly helical configurations form helicity condensates that deviate from these scalings.
Weak turbulence scaling 
was absent from the investigation. 
Finally, the magnetic energy spectra showed support for the Kolmogorov spectrum $k^{-5/3}$ while kinetic energy spectra are closer to
the Iroshnikov-Kraichnan spectrum $k^{-3/2}$. 
\end{abstract} 

\maketitle




One of the most fundamental questions that can be asked about an out-of equilibrium system is the relation
between the energy injection/dissipation rate $\epsilon$, and the amplitude of the fluctuations $u_\ell$.
In hydrodynamic turbulence such estimates are clear and the desired relation comes from the balance between 
the injection rate and the flux of energy to the small scales due to nonlinear interactions.
Such considerations lead to the turbulent scaling 
                      $\epsilon \propto u_{\ell  }^3/\ell  $ for $Re\propto u_{\ell  } \ell/ \nu  \gg 1$.
Here $\nu$ is the kinematic viscosity, $u_\ell$ is the amplitude of the velocity fluctuations at the forcing scale $\ell=\ell_{_F}$.     
Constancy of the energy flux over all scales $\ell$ leads to $u_\ell \propto \epsilon^{1/3}\ell^{1/3}$ that results to the Kolmogorov (K41) 
prediction for the energy spectrum $E(k) =C_{_K} \epsilon^{2/3} k^{-5/3}$ \cite{Kolmogorov1941}.
Although the proportionality coefficient $C_{_K}$ for the turbulent scaling is still debatable in the literature \cite{Goto2009},
the scaling is well supported by both experiments and numerical simulations \cite{Kaneda2003,Sreenivasan1995}.

The situation becomes more complex when a linear wave term is introduced in the dynamical equations.
Such terms introduce new timescales in the system and the validity of the previously derived  relations is  in question.
If the amplitude of the wave frequency  $\omega$ 
is sufficiently larger than the amplitude of the fluctuation shear $u_\ell/\ell$ then
the system can be treated within the frame work of weak turbulence theory \cite{Nazarenko_book2011}. Then depending on the
wave resonances (three-wave, four-wave etc) the energy injection rate is decreased and becomes
\beq
        \epsilon \propto C \, \frac{u_{\ell}^3 }{ \ell} \, \left(\frac{u_{\ell}/\ell}{\omega} \right)^n 
\label{WT}
\eeq
where  $n=0$ for strong turbulence and $n=1$ for three-wave interactions \cite{Nazarenko_book2011}. 
As the strength of the wave term is increased a transition
is expected  from the Kolmogorov scaling for which the energy dissipation rate scales like 
the fluctuation amplitude to the cubic power to the wave turbulent scaling where higher powers are involved.

Magneto-Hydro-Dynamic (MHD) turbulence is an example of turbulence where eddies and (Alfven) waves coexist \cite{Zhou2004,Biskamp2003},
and provides a good approximation to a variety of industrial and astrophysical flows \cite{Zeldo1990,Davidson2001}.
When a flow is coupled to a uniform magnetic field $B_0$ fluctuations travel parallel or anti-parallel to the magnetic field lines with 
$\omega \propto \pm B_0/\ell_\|$. The indexes $\perp,\|$ indicate the direction  perpendicular and parallel to the magnetic field respectively.
The nonlinearities allow for three wave interactions  \cite{Galtier2000} and
the scaling  $\epsilon = C u_{\ell}^4/B_0\ell$ is expected. If isotropy is assumed ($\ell_\perp\sim\ell_\|\sim\ell$),
constant energy flux over all scales $\ell$ implies
$u_{\ell} \propto (\epsilon B_0 \ell )^{1/4}$ that leads to Iroshnikov-Kraichnan (IK) energy spectrum $E(k)\propto  (\epsilon B_0)^{1/2} k^{-3/2}$ 
\cite{Iroshnikov1963,Kraichnan1965}.

However the assumption of isotropy that was assumed is not valid in the presence of a strong magnetic field that suppresses the cascade in its direction \cite{Alexakis2007}. 
For strong turbulence K41 arguments can be repeated 
taking into account anisotropy that leads to the energy spectrum $E(k) \propto \epsilon^{2/3} k_\perp^{-5/3}$  with the so called critical balance relation
                                                   $u_\ell / \ell_\perp \sim B_0 /\ell_\| $
\cite{Goldreich1995}.
%
An anisotropic  IK energy spectrum can also be recovered $E(k) \propto \epsilon^{2/3} k_\perp^{-2/3}$
by assuming that the proportionality coefficient $C$ that appears in eq. \ref{WT} decreases with scale like $ C \propto \ell_\perp^{1/4}$
due to scale dependent alignment of the two fields (velocity and magnetic) that quenches the non-linearities \cite{Boldyrev2006}.
The two proposed spectra have been investigated, in the last years by numerical simulations \cite{Beresnyak2011,Mason2012}
by different groups without however reaching agreement for the strong turbulence spectrum.

Weak turbulence on the other hand valid for $u_\ell / \ell_\perp\ll B_0 /\ell_\| $ can be treated perturbatively.
Detailed calculation \cite{Galtier2000} leads to the prediction for the spectrum $E(k)\propto  f(k_\|) k_\|^{1/2} (\epsilon B_0)^{1/2} k_\perp^{-2}$.
However weak MHD turbulence has the particularity that all three-wave resonant interactions contain one wavenumber
with zero projection in the direction of the magnetic field and zero phase velocity.
The set of these wave numbers themselves compose a strongly coupled nonlinear system and thus further assumptions are needed for the validity
of weak turbulence. In numerical simulations weak MHD turbulence has been observed only when
the resonant manifold $k_\|=0$ is not forced \cite{Perez2007,Bigot2011,Alexakis2011}.
When it is forced the 2D components dominate, the turbulence is strong
and can also introduce an inverse cascade as observed in 2-dimensional flows \cite{Alexakis2011}.

In nature uniform magnetic fields do not exist. However, magnetic fields $B_{L}$ that vary over large length-scales $L$
can be approximated as uniform provided that turbulent energy remains in much smaller scales.
The validity of this approximation however is in doubt since 
small scale variations $\ell_\perp \ll L$ couple to 
parallel variations $\ell_\| \sim B_{L} \ell_\perp /u_\ell$ that
can be as large as $L$ provided that $B_{L}$ is strong enough.
Then the assumption of uniformity is not valid.
Thus, MHD turbulence in the presence of large scale (but not uniform) fields needs to be revisited.
  
%
%
%

To study statistically isotropic MHD turbulence in the presence of large scale magnetic fields
we employ high resolution direct numerical simulations of the MHD equations:
\begin{eqnarray}
\partial_t {\bf u} + {\bf u \cdot \nabla u} & =& {\bf b \cdot \nabla b} -\nabla P +\nu  \nabla^2 {\bf u} +               {\bf F}_u     \label{NS3}\\
\partial_t {\bf b} + {\bf u \cdot \nabla b} & =& {\bf b \cdot \nabla u}           +\eta \nabla^2 {\bf b} +               {\bf F}_b.    \label{MF3}
\end{eqnarray}
in a triple periodic box of size $L=2\pi$.
Here ${\bf u}$ is the velocity field and ${\bf b}$ the magnetic field. 
Both fields satisfy $\nabla \cdot {\bf u} = \nabla \cdot {\bf b} =0$ and 
$\langle {\bf u} \rangle = 
 \langle {\bf b} \rangle =0$,
where the angular brackets stand for spatial average.
$\nu$ is the viscosity and $\eta$ the  magnetic diffusivity.
$F_u$ is an external mechanical body force, while ${\bf F}_b=\nabla \times \mathcal{E}$ where $\mathcal{E}$ is an external electromotive force. 
${\bf F}_u$ and ${\bf F}_b$ are both solenoidal functions varying randomly in time with time correlation $\tau$. 
${\bf F}_u$ is acting only on wavenumbers with $|{\bf k}|=k_u=2$ and is non-helical:
$\langle {\bf F}_u {\bf \cdot \nabla \times F}_u\rangle=0$.
${\bf F}_b$ is acting only in the largest scale of the system $|{\bf k}|=k_b=1$, and in general has non-zero helicity.
All the parameters of the runs can be found in table I. 
For the simulations a pseudo-spectral code was used \cite{Minini_code1,Minini_code2} on grids of size $512^3$ (Runs A\#) and $1024^3$ (Runs B\#).
%

\begin{figure}                                                                                                                %
\includegraphics[width=8.0cm]{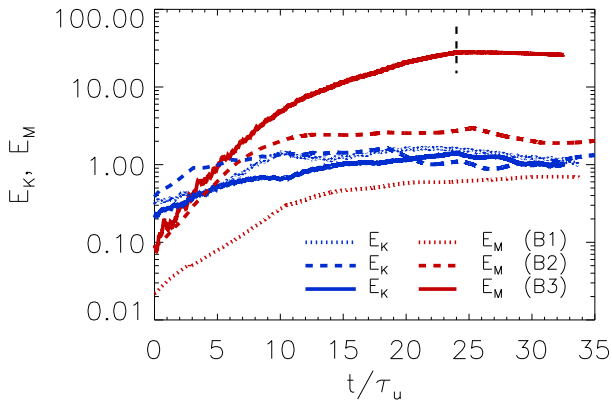}                                                                                     %
\caption{\label{fig1} Kinetic and magnetic energy evolution as a function of time for the three $1024^3$ runs. The vertical   %
dashed line indicates when the forcing parameters changed from the ones for B3' to the ones for B3.}                          %
\end{figure}

In a typical helical run 
all quantities grow initially
up to a point when dissipation rates and kinetic energy reach a steady state while the magnetic energy is still increasing slowly. 
During this time the magnetic field is composed of a large scale helical component $B_L$ with $|{\bf k}| \simeq 1$ that contains most of the magnetic energy
and small scale turbulent fluctuations $b_\ell$ of amplitude $b_\ell \sim u$. 
Thus magnetic energy $E_{_M}$ provides a measure of the large scale field $E_{_M}\simeq \frac{1}{2} B_{L}^2$, while kinetic energy $E_{_K}$
provides a measure of the turbulent fluctuations.
The growth of $B_{L}$ depends on the amplitude and the helicity of the magnetic forcing.
%
%
The time evolution of the large resolution runs can be seen in figure \ref{fig1}. 
Due to the slow increase of the magnetic energy it is possible to perform short time averages (over a few turn over times) and thus obtain global averaged 
quantities for various values of magnetic energy from the same run. Thus the turbulent scaling of the energy dissipation rate can be tested from multiple 
measurements for a wide range of magnetic field strength that here it is quantified by $\mu\equiv E_{_M}/E_{_K}$. The isotropy of the system also allows to
to perform spherical averages and thus improve two point statistics compared to the case with uniform magnetic fields.


\begin{table}[ht]                                                                                                                                                               %
\label{tbl}                                                                                                                                                                     %
\begin{tabular*}{0.45\textwidth}{@{\extracolsep{\fill}}                                                                                                                         %
               | c        ||     c               |      c          |   c          |   c                ||   c            | c                    |  c         | }                %
\hline                                                                                                                                                                          %
                 RUNS     & $\mathcal{G}^{1/2}$  & $\mathcal{M}$ & $\mathcal{H}\,$&$\mathcal{S}t^{-1}$ & $\mathcal{R}e$ & $\mathcal{R}_\lambda$ &  $\mu$        \\              %
\hline \hline             
                    A1    &   356                &     0.00        &      -       &   0.50             &   580          &     157               &     0.4       \\              %
                    A2    &  1666                &     0.10        &     0.59     &   0.01             &   905          &     181               &     0.7       \\              %
                    A3    &  2500                &     0.00        &      -       &   0.01             &  1331          &     206               &     0.7       \\              %
                    A4    &  1666                &     1.00        &     0.00     &   0.01             &  1064          &     170               &   1.4 - 1.7   \\              %
                    A5    &   353                &     0.50        &     0.15     &   0.50             &   541          &     106               &   1.5 - 2.9   \\              %
                    A6    &   353                &     0.50        &     0.31     &   0.50             &   690          &     151               &   2.4 - 3.2   \\              %
                    A7    &   353                &     0.50        &     1.00     &   0.50             &   734          &     205               &   2.6 - 4.5   \\              %
                    A8    &   353                &     0.50        &     0.95     &   0.50             &   645          &     148               &   2.7 - 5.0   \\              %
                    A9    &   353                &     1.00        &     0.15     &   0.50             &   811          &     141               &   1.8 - 7.6   \\              %
                    A10   &  1666                &     1.00        &     1.00     &   0.01             &   972          &     154               &   1.8 - 9.5   \\              %
                    A11   &  1666                &     1.00        &     0.59     &   0.01             &   920          &     138               &   2.0 - 14.0  \\              %
                    A12   &  1666                &     1.00        &     1.00     &   0.01             &  1877          &     634               &   5.7 - 12.0  \\              %
                    A13   &  1054                &    10.00        &     0.59     &   0.01             &  1081          &     105               &   6.0 - 94.0  \\\hline \hline %
                    B1    &   3846               &     0.00        &      -       &   0.01             &   2813         &     296               &    0.6        \\              %
                    B2    &   1414               &     0.50        &     0.31     &   0.5              &   2572         &     311               &    2.6        \\              %
                    B3$'$ &   6123               &     1.00        &     1.00     &   0.01             &   3714         &     386               &  5.0 - 18.0   \\              %
                    B3    &   6123               &     0.40        &     0.50     &   0.01             &   4105         &     298               & 18.0 - 24.0   \\ \hline       %
\end{tabular*}                                                                                                                                                                  %
\caption{Table with the parameters of all runs. The numbers in the first four columns are the input parameters given by                                                         %
$\mathcal{G}            \equiv \langle {\bf F_u}^2 \rangle^{\frac{1}{2}} / \nu^2 k_u^3$ the Grashof number,                                                                     %
$\mathcal{M}            \equiv \langle {\bf F_b}^2 \rangle^{\frac{1}{2}} /\langle{\bf F_u}^2\rangle^{\frac{1}{2}}$ the ratio of magnetic to mechanical forcing,                 %
$\mathcal{H}            \equiv \langle {\bf F_b \cdot \nabla \times F_b} \rangle  /\langle{\bf F_b}^2\rangle k_b$ the relative helicity of the forcing and                      %
$\mathcal{S}t           \equiv \tau_f / \tau$ the Strouhal number where                                                                                                         %
$\tau_f                 \equiv 1 / (\langle {\bf F_u}^2\rangle k_u^2)^{1/4}$.                                                                                                   %
$\mathcal{M}=0$ imply dynamo runs.                                                                                                                                              %
The Prandtl number $\mathcal{P}r \equiv \nu/\eta$ for all runs was set equal to 1.                                                                                              %
The last columns give the measured parameters                                                                                                                                   %
$\mathcal{R}e           \equiv \langle{\bf u}^2\rangle^{\frac{1}{2}}   /\nu            k_u              $ the        Reynolds number and                                        %
$\mathcal{R}_\lambda   \equiv \langle{\bf u}^2\rangle^{\frac{1}{2}} /\nu  \langle({\bf \nabla \times u})^2\rangle^{\frac{1}{2}}  $ the Taylor Reynolds number.                                                                      %
$\mathcal{R}_\lambda    \equiv \langle{\bf u}^2\rangle^{\frac{1}{2}} \lambda /\nu $ (with $\lambda^2 \equiv \langle{\bf u}^2\rangle/\langle(\bf \nabla \times u)^2\rangle $)      %
the Taylor Reynolds number.                                                                                                                                                     %
The last column gives the range of values obtained by                                                                                                                           %
$\mu\equiv E_{_M}/E_{_K}$ the ratio of magnetic to kinetic energy.                                                                                                              %
The simulations named $A$\# were carried out in a grid of size $512^3$ while the simulations named $B$\# were carried out in a grid of size $1024^3$.                           %
The third $1024^3$-grid simulation started with the parameters of B3$'$ and run up to $t \simeq 24$ turn over times $\tau_u\equiv 1/k_u \langle{\bf u}^2\rangle^{\frac{1}{2}}$. %
afterwards it was continued wit the parameters B3.                                                                                                                              %
}                                                                                                                                                                               %
\end{table}                                                                                                                                                                     %

\begin{figure}                                                                                                                %
\includegraphics[width=8.0cm]{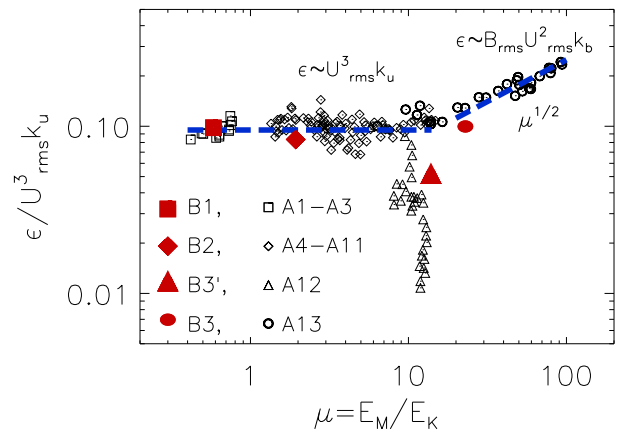} \\                                                                                  %
\caption{\label{fig2}                                                                                                         %
Top    panel: The energy dissipation rate $\epsilon$ normalized by $U_{rms}^3k_u$ as a function of $\mu=E_{_M}/E_{_K}$        %
              for all runs. $B_{rms}=\sqrt{2E_{_M}}$ and $U_{rms}=\sqrt{2E_{_M}}$.                                            %
              The dashed lines indicate the scaling $\epsilon\propto U_{rms}^3k_u$ and                                        %
              $\epsilon  \propto U_{rms}^2 B_{rms} k_b \propto \mu^{1/2}$.                                                    %
}                                                                                                                             %
\end{figure}

Figure \ref{fig2} shows the energy dissipation rate normalized by $U_{rms}^3k_u= (2E_{_K})^{3/2}k_u$
as a function of the energy ratio $\mu$ for all the examined runs. The data cover more than two decades of the parameter $\mu$. 
Three different behaviors can be observed.
First,
over the range $\mu$ (0.5$\le\mu\le$20) the energy dissipation is independent of $\mu$.
This implies that $\epsilon$ follows the Kolmogorov scaling $\epsilon \propto u_\ell^3/\ell$ 
even when the large scale magnetic energy is twenty times greater than the turbulent kinetic energy.
The data include runs that vary from fully helical to non-helical, and strongly magnetically forced
to dynamo runs and for $\mathcal{R}_\lambda\sim 100$ to $\mathcal{R}_\lambda\sim 300$.
Thus this result seems to be very general and robust in this range.

At $\mu$ larger than 10 two new branches appear.
The results of run $A12$ that is fully helical and strongly magnetically forced is marked by triangles in figure \ref{fig2}.
For this run both magnetic and kinetic energy is concentrated in the large scales building
helical structures with very small turbulent fluctuations
The large scale magnetic and kinetic energy both increase with time keeping their ratio $\mu$ fixed
while the small scale fluctuations and the dissipation rates saturate with time.
As a result 
the normalized dissipation rate $\epsilon/U_{rms}^3k_u$ decreases with time resulting in the behavior seen in fig.\ref{fig2}. 
The dynamics here are controlled by magnetic helicity condensates, and despite the large
Reynolds number they do not follow a turbulent scaling.
 
Run $A13$ marked by circles in figure \ref{fig2} is strongly magnetically forced in order to achieve large values of $\mu$ within the
time limitations imposed by the computational costs. This run although in agreement with the turbulent scaling for $\mu\le20$
it transitions to the scaling $\epsilon\propto \mu^{1/2}$ as indicated by the dashed line in the figure. This scaling can be understood 
if we consider that the main mechanism for cascading the injected energy is not the velocity shear $S_u \propto U_{rms}k_u$ but rather
the magnetic shear $S_b\propto B_{rms}k_b$ that shreds Alfven-wavepackets as they travel along chaotic magnetic field lines.
The resulting scaling is $\epsilon\propto S_b U^2 \propto U^3k_b \mu^{1/2}$ that is observed in figure \ref{fig2}. 
Thus the large scale field rather than suppressing the turbulence cascade it enhances it.

None of the runs showed a weak turbulence scaling that would have implied according to \ref{WT} the scaling
$\epsilon\propto U^4/B_{L}k_u \propto \mu^{-1/2}$. There are few possible interpretations for this result. 
%
First,  just like the case of the uniform magnetic field, the absence of weak turbulence is can be explained by the entrapment of energy
in modes that vary perpendicular to the local magnetic field so that $\ell_\perp\ll\ell_\|$ that makes the nonlinear coupling strong. 
Another possibility is the lack of uniformity of the magnetic energy density:
regions exist in space with weak local magnetic field where eddies can be stretched 
with no resistance from magnetic tension. If the cascade in these regions dominates
the scaling of turbulence becomes independent of $B_{L}$.
%
A final possibility is that magnetic instabilities reveal themselves altering the scaling (like run A13) before the magnetic field
becomes strong enough for weak turbulence to manifest itself. 


%

\begin{figure}                                                                                                                %
\includegraphics[width=8.0cm]{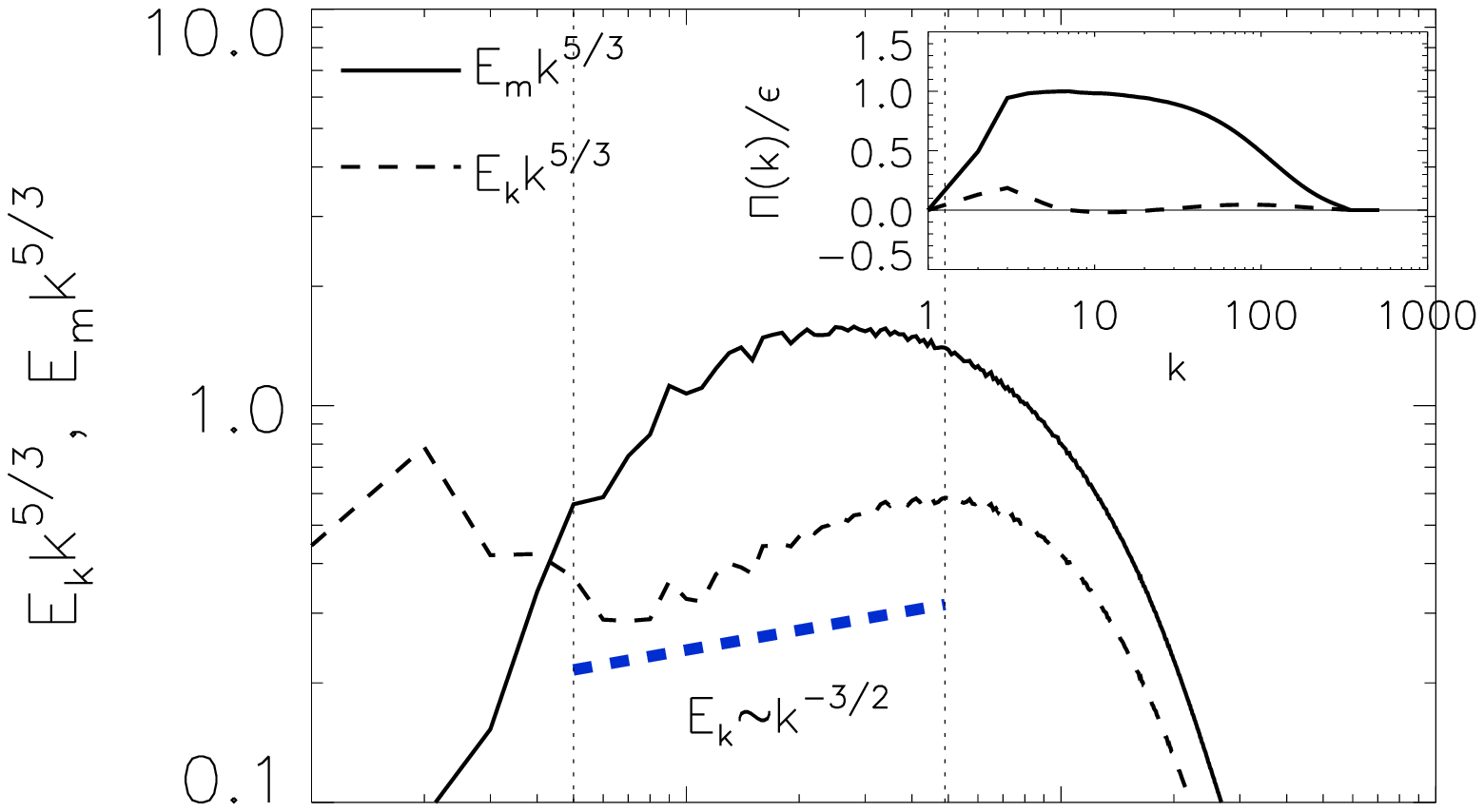}                               \hspace{0.0cm}    \\                                  %
\includegraphics[width=8.0cm]{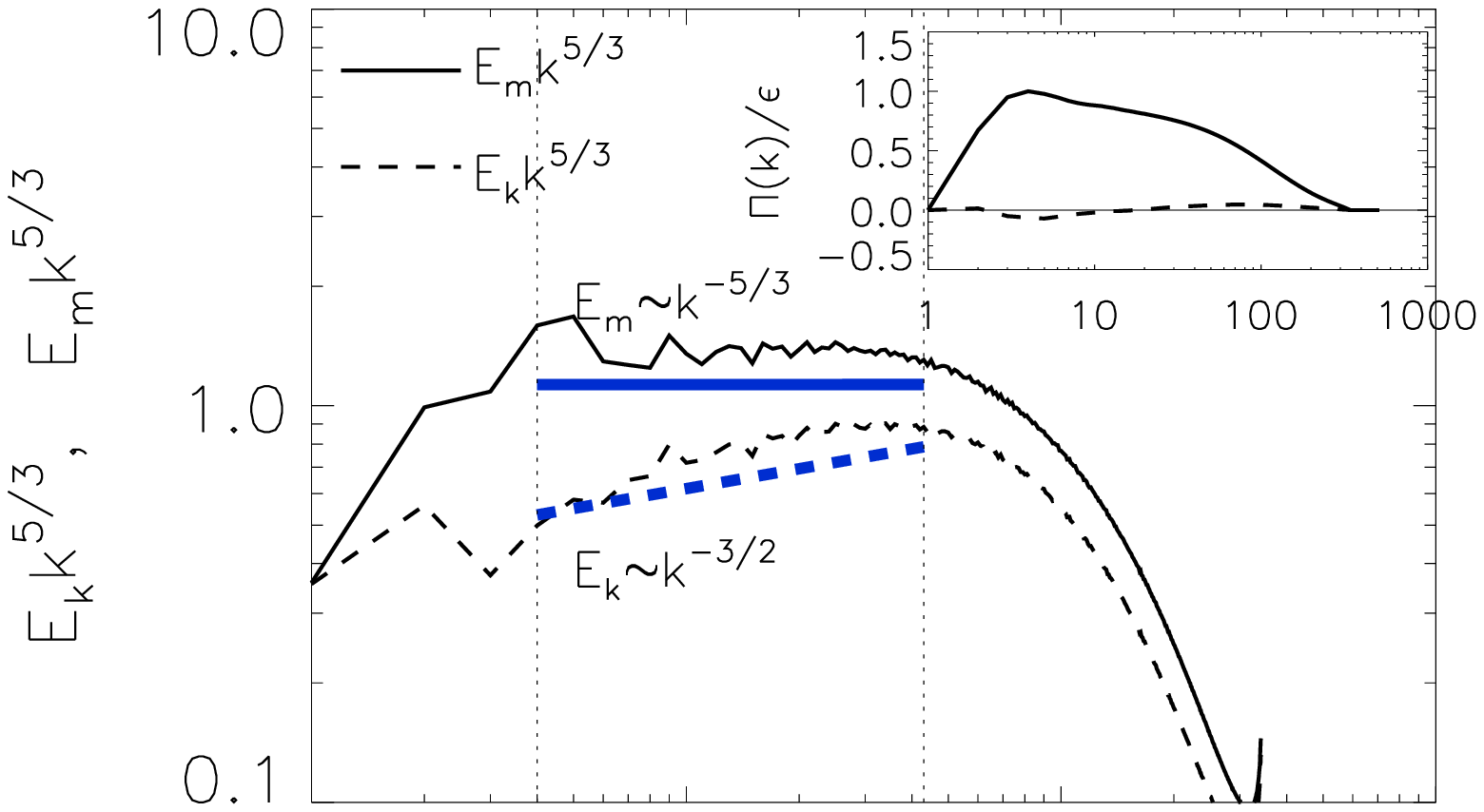}                               \hspace{0.0cm}    \\                                  %
\hspace{0.05cm} \includegraphics[width=8.2cm]{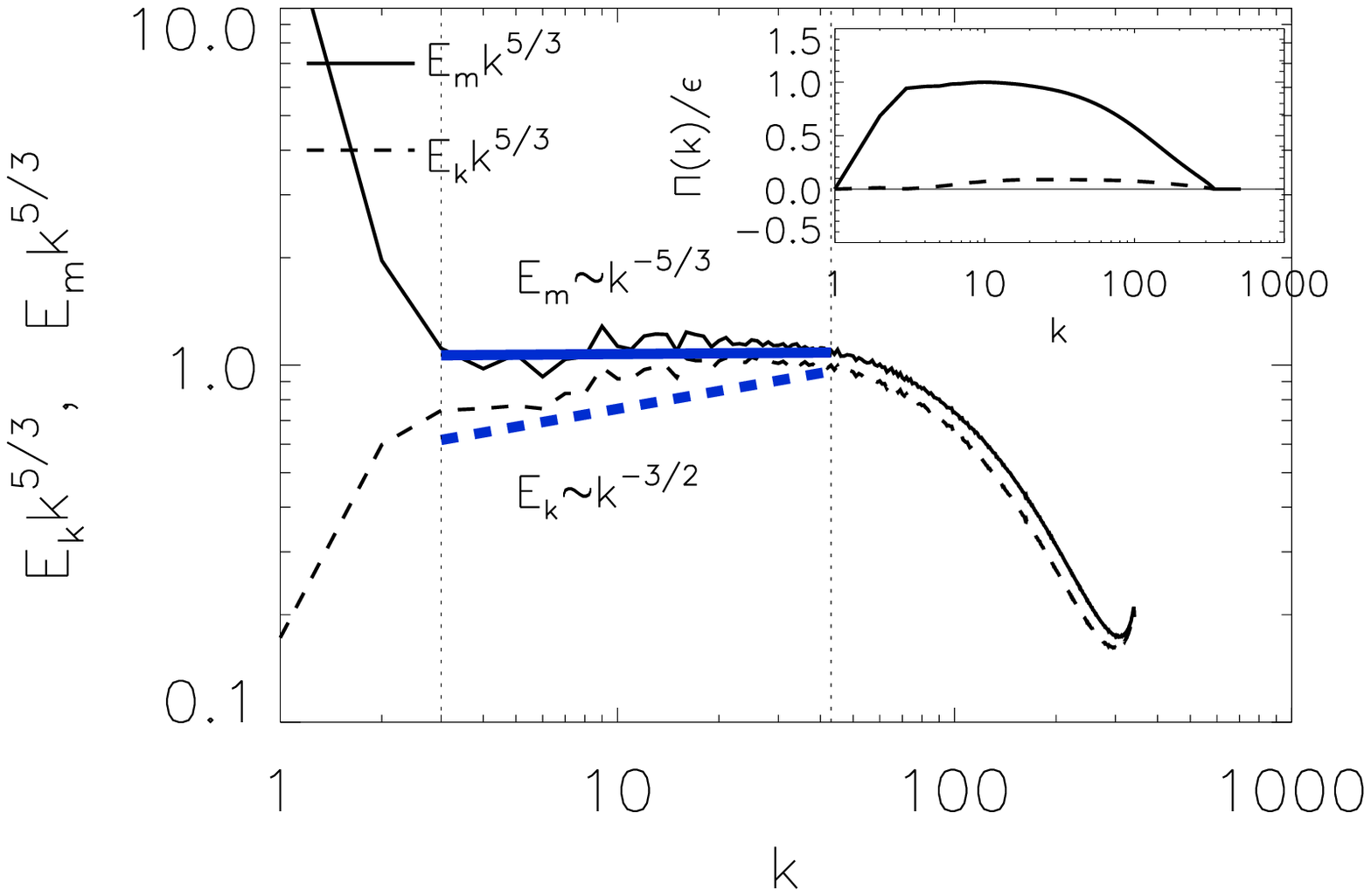}                                                                     %
\caption{\label{fig3}                                                                                                         %
Kinetic and magnetic compensated energy spectra for the Runs B1,B2,B3 (from top to bottom). The straight lines indicates the  %
scaling $k^{-5/3}$ (solid) and $k^{-3/2}$ (dashed). The inset shows the energy flux $\Pi(k)$ (solid line)                     %
and $\Pi_u(k)$ (dashed line).                                                                                                 %
}                                                                                                                             %
\end{figure}

The energy spectra $E_m,E_k$ for the high resolution runs B1,B2,B3 are shown in figure \ref{fig3}. 
For the B1 dynamo run the kinetic energy spectrum shows a power-law scaling slightly less steep than the $k^{-3/2}$ prediction.
A clear change of slope is observed at $k\sim8$ where $E_m(k)$ becomes larger than  $E_k(k)$ is also observed.
The magnetic energy spectrum does not show  a power-law scaling.
The absence of a power-law scaling might be linked to the absence of large scale magnetic structures in randomly forced dynamo runs. 
The B2, B3 runs despite the one order of magnitude difference in $\mu$ show similar results. The kinetic energy
spectrum is well fitted by a $k^{-3/2}$ power-law while the magnetic energy spectrum is best fitted by a $k^{-5/3}$ law. 
%
%
The fact that different exponent is measured for the velocity and for the magnetic field
indicates that probably a clear inertial range has not been reached yet even at $1024^3$ grid sizes. 
Two reasons affect the measured  spectrum exponents. First, at small wave numbers the large scale field $B_L$
contaminates the inertial range making the spectrum look steeper. Second at large wave numbers the energy 
spectrum can be altered by bottle-neck effects that are far from well understood in MHD.
%
The insets in figure \ref{fig3} show the total energy flux 
$\Pi(q) \equiv \langle {\bf  u}_q^< {\bf   ( u\cdot\nabla u  - b\cdot\nabla b)  +  b}_q^< {\bf ( u\cdot\nabla b - b\cdot\nabla u)}\rangle $ 
\cite{Biskamp2003} (solid lines) 
and the energy flux due to the vortex stretching term $\Pi_u(q) \equiv \langle {\bf  u}_q^< {\bf (u\cdot\nabla u)} \rangle$ (dashed line).
${\bf u}_q^<,\, {\bf b}_q^<$ are the filtered velocity field containing only wavenumbers with $|{\bf k}|<q$. 
In all three cases the vortex stretching term plays a minor role in cascading the energy.

\begin{figure}                                                                                                                
\vspace{0.2cm}                                                                                                                
\includegraphics[width=8.0cm]{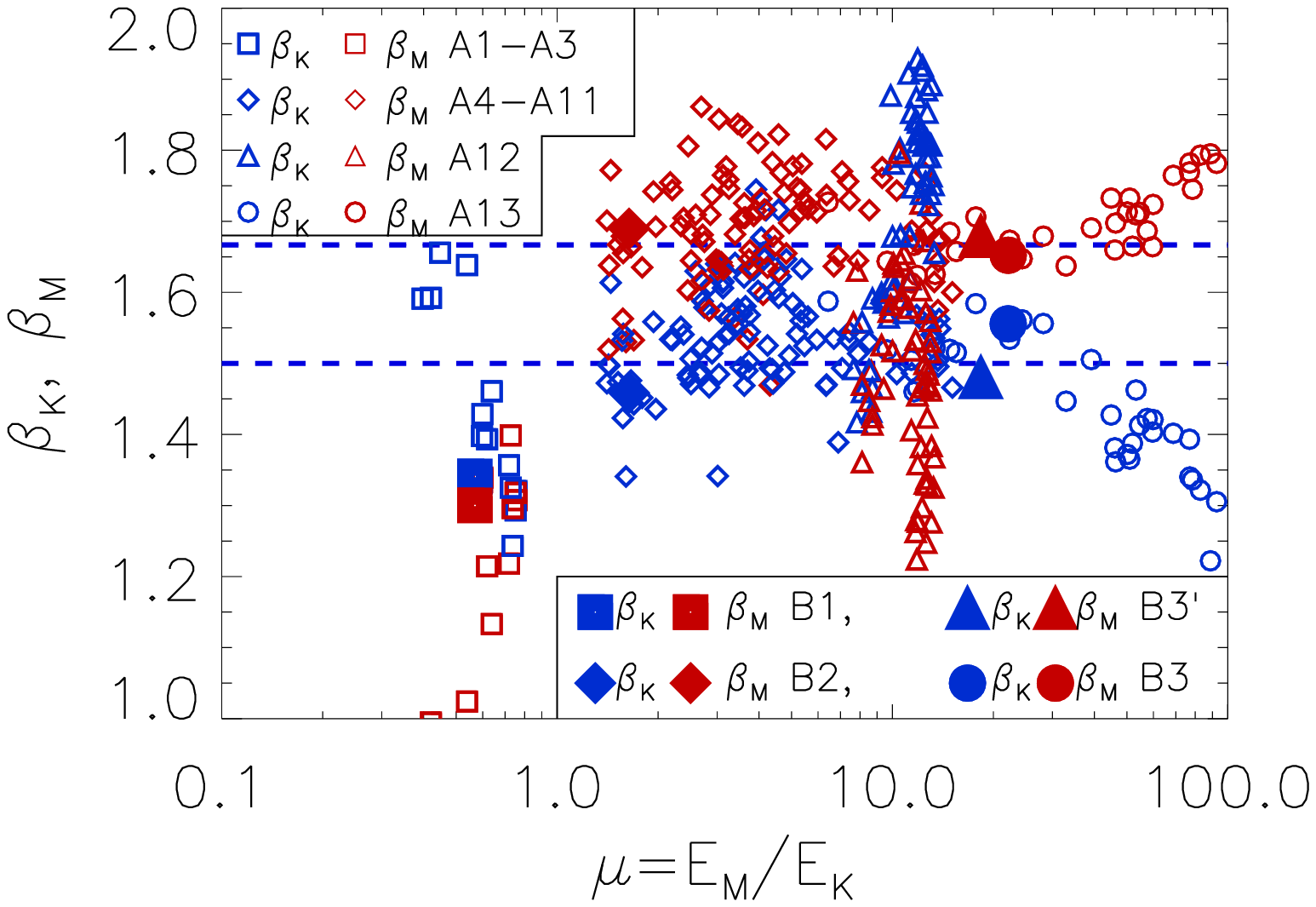} \\                                                                                  
\includegraphics[width=7.6cm]{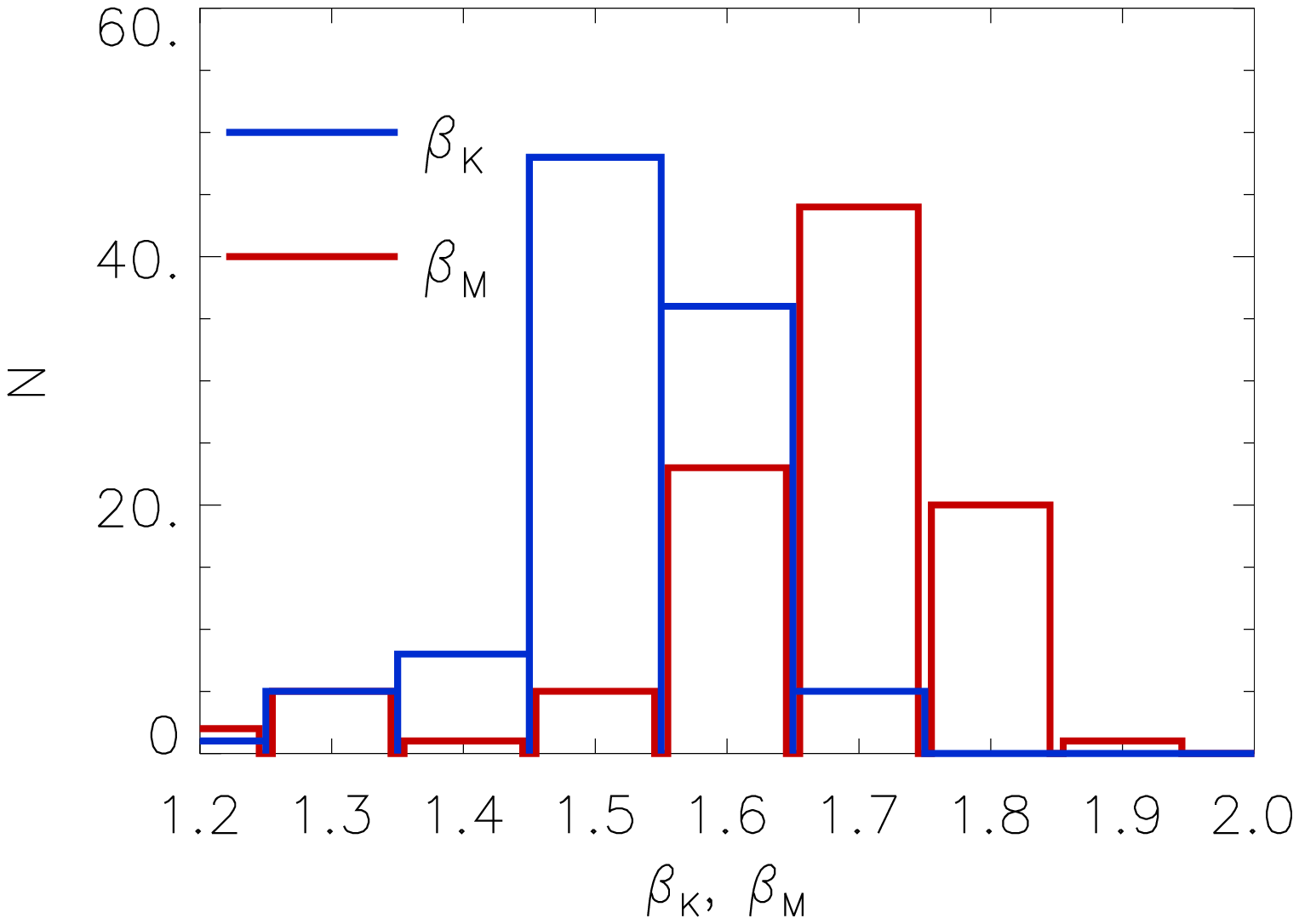}                                                                                     
\caption{\label{fig4}                                                                                                         
Top    panel: Measured exponents $\beta_{_K}$ and $\beta_{_M}$  for all runs. The exponents were calculated by a linear fit   
in the range $2k_u < k <\frac{3}{4} k_{\nu}$, where $k_{nu}$ corresponds to the peak of the energy spectrum.                  
Bottom panel: Distribution of the measured exponents for all runs in the range $1<E_{_M}/E_{_K}<20$, excluding run A12        
and A1-A3.    }                                                                                                               
\end{figure}                                                                                                                  


The top panel of figure \ref{fig4} shows the measured power-law exponents from all the runs. The exponents were calculated by 
fitting power-law solutions $E_k \sim A_k k^{-\beta_K}$ and $E_m \sim A_m k^{-\beta_M}$ 
in the range $2k_u < k < \frac{3}{4}  k_{\nu}$, where $k_{\nu}$ corresponds to the peak of the enstrophy spectrum $k^2E_k$. 
The range of fitting can be seen in fig.\ref{fig3} by the vertical dotted lines.
The bottom panel shows the distribution of these exponents in the range $1<\mu<20$,           
excluding run A12 and the A1-A3 dynamo runs.
Although the dispersion of these values is quite large the exponents do not show much dependence on amplitude of the magnetic field.
Their values are concentrated around $k^{-5/3}$ for the magnetic energy spectrum and $k^{-3/2}$ for the kinetic energy spectrum
in agreement with the high resolution runs.

Concluding, this work has showed an independence of the normalized energy dissipation rate ($\epsilon\propto U_{rms}^3k_u$)
and spectral exponents ($\beta_{_M}\simeq 5/3$ and $\beta_{_K}\simeq 3/2$)
on the amplitude of the large scale magnetic field over a large range of $\mu$ and for a variety of forcing configurations.
%
Nonetheless, deviations were observed for very strong magnetic fields $\mu>20$ (for which ($\epsilon\propto B_{rms}U_{rms}^2k_b$)
and for fully helical flows that formed magnetic helicity condensates in the large scales.   
%
%
Analysis of higher order moments \cite{Biskamp2003} and scale interactions \cite{Alexakis2005} might shed more light in the processes that control MHD turbulence.




{\it Aknowledgements. }
This work was performed using HPC resources from  GENCI-CINES
(Grant 2012026421).




%

\end{document}